\documentclass[dvips]{article}
\usepackage{icrctc07}

\title{Synchrotron Emissions in GRB Prompt Phase Using a Semi \\
Leptonic and Hadronic Model}
\shorttitle{Synchrotron emissions in GRBs}
\authors{S. Guiriec$^{1}$, D. Gialis$^{2}$, G. Pelletier$^{2}$, F. Piron$^{1}$.}
\shortauthors{S. Guiriec et al.}
\afiliations{$^1$ LPTA, Universit\'e Montpellier 2, CNRS/IN2P3, Montpellier, France\\ $^2$ LAOG, Universit\'e Joseph Fourier, CNRS/INSU, Grenoble, France}
\email{sylvain.guiriec@lpta.in2p3.fr}

\abstract{In this communication devoted to the prompt emission of GRBs, we claim that some important parameters associated to the magnetic field,
such as its index profile, the index of its turbulence spectrum and its level of irregularities, will be measurable with GLAST.
In particular the law relating the peak energy $E_\mathrm{peak}$ with the total energy $E$ (like Amati's law) constrains the turbulence spectrum
index and, among all existing theories of MHD turbulence, is compatible with the Kolmogorov scaling only.
Thus, these data will allow a much better determination of the performances of GRBs as particle accelerators. 
This opens the possibility to characterize both electron and proton acceleration more seriously.
We discuss the possible generation of UHECRs and of its signature through GeV--TeV synchrotron emission.}

\begin{document}
\maketitle

\section{Introduction}
Future GLAST observations of GRBs, especially of their prompt emission, will considerably improve
our knowledge of GRB phenomena and our diagnosis of their high-energy performances.
In the following we will show that better constraints of the main magnetic parameters are indeed expected, thus a better knowledge of the
acceleration conditions for both electrons and protons.

Our investigation concerns the prompt stage of GRBs interpreted in terms of multiple internal
shocks~\cite{ref7}.
The parameter that we consider as less constrained is the intensity of the field at some point,
either at the origin scale $r_0$ of a few gra\-vi\-ta\-tio\-nal radii, or at the beginning of the ac\-ce\-le\-ra\-tion stage in the internal shocks
at $r_b = \eta^2 r_0$, $\eta$ being the baryonic parameter $(E/M_bc^2)$.
How\-ever we think that the profile of the mean field is already more or less constrained and will be better constrained soon,
namely the index $\alpha$ such that $B \propto r^{-\alpha}$ can be determined.
The other important index that controls the particle acceleration efficiency as a function of the particle energy
is the index of the turbulent spectrum $\beta$ ($\beta= 5/3$ for Kolmogorov theory). The efficiency of the Fermi
acceleration process is directly linked with the efficiency of particle scattering off magnetic ir\-re\-gu\-la\-ri\-ties.
The most efficient acceleration is obtained with the so-called Bohm scaling, which corresponds to $\beta=1$.
Of course this efficiency is also proportional to the level of magnetic irregularities $\eta_t\equiv$ $<$$\delta B^2$$>$$/$$<$$B^2$$>$.
We claim that these three parameters $\alpha$, $\beta$ and $\eta_t$ will be determined quite nicely by GLAST campaigns
on prompt stage of GRBs. The reason is that the position of $E_\mathrm{peak}$,
its evolution with time, and more generally the evolution of the synchrotron spectrum,
moreover the value of the maximum synchrotron emitted ener\-gy all simply depend on these three parameters.
The relation between $E_\mathrm{peak}$ and the fireball ener\-gy $E$, in the spirit of Amati~\cite{ref1},
maybe not tight enough for cosmological purpose, leads to a surprising constraint on $\beta$, as will be seen later on.
\begin{figure}[t!]
\begin{center}
\includegraphics [width=0.4\textwidth]{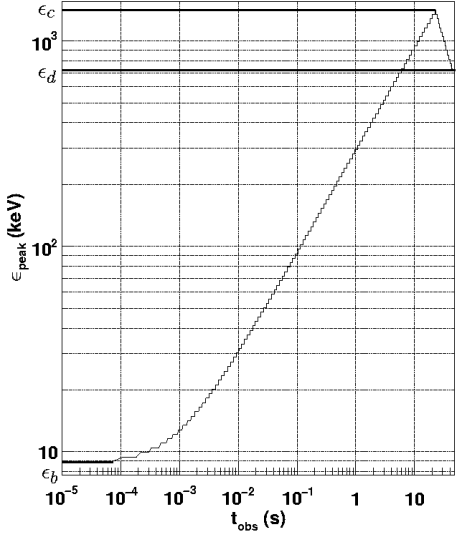}
\end{center}
\caption{{\footnotesize Evolution of $\epsilon_\mathrm{peak}$ in the observer frame. The instantaneous value of $\epsilon_\mathrm{peak}$
increases from the broa\-de\-ning radius $r_b$ to the radius $r_c$ where the dominant coo\-ling changes and then decreases down to the
deceleration radius $r_d$. Depending on the magnetic parameter $\alpha$ (here $\alpha$=1),
this $\epsilon_\mathrm{peak}$ evolution can be fully observed or not with a gamma-ray telescope.}}\label{fig1}
\end{figure}

Then we can derive the condition for Fermi ac\-ce\-le\-ra\-tion of relativistic electrons and make a more stringent
prediction on cosmic-ray generation. Recently a detailed analysis~\cite{ref2} of cosmic-ray ac\-ce\-le\-ra\-tion at relativistic
shocks emphasized the necessity of generating an intense turbulence at short scale in order to get several Fermi cycles.
This problem received a solution that is in course of publication~\cite{ref3}, but the possibility of generating UHECRs at
the external shock seems hopeless.
Thus the regime of internal shocks seems to be the main possibility of getting UHECRs so far~\cite{ref4}.
However, we will see that the usual Fermi acceleration at shocks is not sufficient to reach the UHECR energies
and that a secondary acceleration by scattering off the magnetic fronts themselves is needed.
These UHECRs radiate a synchrotron emission in the GeV--TeV range and the ob\-ser\-va\-bi\-li\-ty of that emission will be discussed.

\section{Determination of magnetic parameters}
\begin{figure}[t!]
\begin{center}
\includegraphics [width=0.4\textwidth]{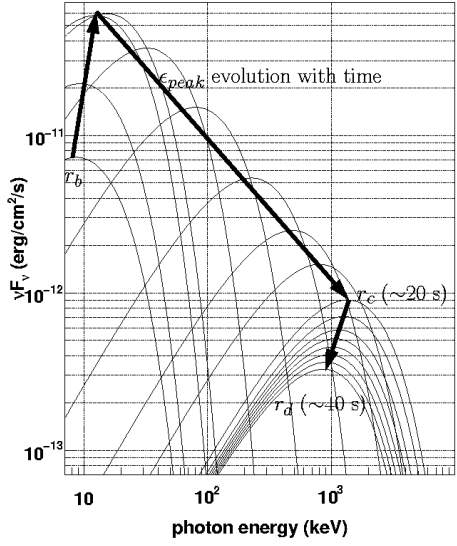}
\end{center}
\caption{\footnotesize Evolution of the synthetic instant spectrum from the electron synchrotron emission, in the observer frame.
The following magnetic parameters have been used: $B(r_b)$$=$$10^5$~G, $\alpha$$=$$1$, $\beta$$=$$5/3$~(Kolmogorov), $\eta_t$$=$$1.4\times10^{-2}$. 
The instant $\epsilon_\mathrm{peak}$ value is increasing during the first $20$~s (i.e. till $r_c$) from a few keV up to $\sim$5~MeV
(Fig.~\ref{fig1}), then decreasing down to hundreds of keV (i.e. till $r_d$).
The full shock dynamic is not implemented yet in this simulation.
Instead, we increased by hand the fraction of accelerated electrons during the shock phase.
This causes a rise of the flux during the first instants.}\label{fig2}
\end{figure}

The electron energy distribution undergoes a two stage self-similar evolution. During the first stage from $r_b$ to some
distance $r_c$, that depends on magnetic parameters, the electron acceleration is li\-mi\-ted by synchrotron losses.
As the magnetic field decreases like $r^{-\alpha}$, the high energy cut-off of the synchrotron spectrum migrates towards high ener\-gy,
from the highest intensity maximum of ener\-gy $\epsilon_b$ up to some photon ener\-gy $\epsilon_c$ (Fig.~\ref{fig1}~and~\ref{fig2}).
Then the second stage is dominated by expansion coo\-ling and the synchrotron spectrum migrates from $\epsilon_c$ towards lower energies.

Whereas the self-similar evolution of the spectrum during the first stage dominated by synchrotron cooling is governed by $\alpha$
and $\beta$, the self-similar cooling by expansion depends on $\alpha$ only: $\epsilon_{ph} \propto r^{2-3\alpha}$.
This decay law is quite sensitive to the value of $\alpha$ and its observability more likely favors $\alpha=1$.
This indicates the existence of a dominant toroidal component of the magnetic field, which is rea\-so\-na\-bly expected.
As recently pointed out~\cite{ref8}, a se\-cond interesting remark is that the highest cut off, $\epsilon_c \propto \eta \eta_t^{1\over 2-\beta}$,
amazingly depends on the index $\beta$ only and, even more amazingly, is sensitive to the level of magnetic irregularities $\eta_t$
and not to the field intensity. In particular, since we cannot diminish the baryonic parameter $\eta$ below a few tens,
it is quite surprising that the level of turbulence has to be {\it lowered} in order to get a reasonable value of this cut-off! 
Some GRBs should display a short growth of their characteristic energy followed by a decay,
the maximum being observed in the MeV range. This requires a level of turbulence as low as $10^{-3}-10^{-2}$,
depending on $1 \leq \beta < 2$, even for a powerful burst like GRB990123~\cite{ref9}.

Among all the possibilities of constraining $\beta$, the law relating the ``initial" $\epsilon_\mathrm{peak}$, corresponding to the maximum
flux (not the usual one cor\-res\-pon\-ding to an integrated flux) with the fireball energy $E$ (like an Amati's law) is the most interesting:
\begin{eqnarray}
\label{equ1}
\epsilon_\mathrm{peak} \propto \eta^{\gamma \over 3-\beta} \eta_t^2 \left(\frac{\Omega c \Delta t_w}{r_0}\right)^{{3\over 2}{\beta-1 \over 3-\beta}} (M_\mathrm{BH}^{5/2}E_\mathrm{mag}^{-3/2})^{\beta-1 \over 3-\beta} \nonumber
\end{eqnarray}
with $\gamma = 5-3\beta+6\alpha(\beta-1)$.
This $\epsilon_\mathrm{peak}$ is close to $\epsilon_b$, unless the photosphere radius is located much further than $r_b$ (but we will not discuss
these details in this short paper).

It is natural to state that $E_\mathrm{mag}\propto M_\mathrm{BH}\propto E$. Therefore a theoretical extension of the Amati's law is obtained with an index involving $\beta$ only:
\begin{eqnarray}
\label{equ2}
\epsilon_\mathrm{peak} \propto E^{\beta-1 \over 3-\beta}.\nonumber
\end{eqnarray}
The exponent $1/2$ suggested by the observational Amati's law is obtained for the Kolmogorov index $\beta = 5/3$, which is quite
remarkable. No other plausible index like $\beta = 3/2$ (Kraichnan) or $\beta = 2$ is compatible!
A Bohm scaling ($\beta = 1$) is also completely ruled out.
It is a pity that we cannot maintain $\eta_t\simeq 1$, which would considerably reduce the dispersion in Amati's law.
However, if a GRB observation allows to measure $\epsilon_c$ and if an independent constraint on the baryonic
parameter is provided (as for instance a spectral break in the afterglow), then the turbulence level can be estimated and this
dispersion reduced.

\begin{figure}[t!]
\begin{center}
\includegraphics [width=0.40\textwidth]{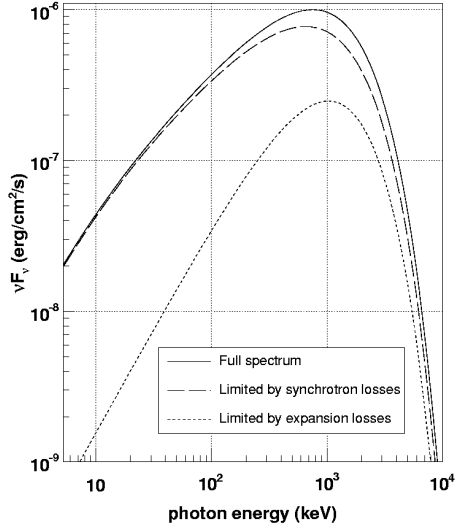}
\end{center}
\caption{\footnotesize Synthetic spectrum of the synchrotron emission of the relativistic electrons, using the same magnetic parameters as in
Fig.~\ref{fig1}~and~\ref{fig2}.
This simulated ${\nu}F{_{\nu}}$ spectrum can be characterized by a Band's function with parameters $E_\mathrm{peak}$$=$$860$~keV,
${\alpha}_\mathrm{Band}$$=$$-0.88$ and ${\beta}_\mathrm{Band}$$=$$-3.1$ (and a total energy of $1.28{\times}10^{51}$~ergs for a $T_{90}$ duration
of $18$~s).
It is similar to the observed spectrum of GRB990123, which has $E_\mathrm{peak}$$=$$720$~keV, ${\alpha}_\mathrm{Band}$$=$$-0.6$ and
${\beta}_\mathrm{Band}$$=$$-3.1$~\cite{ref9}.
}\label{fig3}
\end{figure}

Actually $\alpha=1$, $\beta=5/3$ and a moderate level of turbulence are suitable conditions for fitting the existing data
(Fig.~\ref{fig3}). The maximum electron Lorentz factors achieved by shock acceleration are much more modest than often proposed:
we obtain $\gamma_e\sim 10^2$, instead of $10^5$ that would be incompatible with observations.
An important consequence is that the SSC emission does not reach $10$~GeV. Moreover $L_\mathrm{SSC}/L_\mathrm{syn}\sim10^{-3}$. 
Incidentally, the result by Gonz\'alez et al.~\cite{ref5} of an increasing high-energy spectrum towards GeV energies, without variability
correlated with the synchrotron emission, suggests an hadronic origin. 

\section{Signature of UHECR generation}

For particle acceleration, $\alpha=1$ is very helpful and allows the Hillas criterium to be uniform over the whole range of the internal shocks as expected by Waxman~\cite{ref4}, whereas $\alpha=2$ would limit the acceleration stage to a fairly short distance interval beyond $r_b$.
However, as previously seen in the case of electron acceleration, $\beta=5/3$ instead of $\beta=1$, which makes Fermi acceleration process much less efficient.
For these values of $\alpha$ and $\beta$, one obtains the following energy maximum for the acceleration of protons limited by expansion (measured
in GeV in the co-moving frame):

\begin{eqnarray}\label{equ3}
\epsilon_{exp} \simeq 1.2\times 10^6 \eta_t^3 \left(\frac{\eta}{100}\right)\left(\frac{B(r_b)}{10^5G}\right)
\left(\frac{r_0}{10^7cm}\right).\nonumber
\end{eqnarray}
Even with a strong turbulence level, this estimate shows that UHECRs cannot be produced since the maximum energy measured by an observer would be smaller than $10^{18}$~eV.

A secondary acceleration process is required. It has been proposed~\cite{ref6} that protons accelerated at internal shocks can be scattered off multiple magnetized fronts, the internal shocks themselves, and undergo a kind of ``second order" Fermi process, which is efficient in this mildly relativistic regime.
At each scattering the cosmic rays have an ener\-gy gain of order $\gamma_*^2$, where $\gamma_*$$\sim$$2$ is the average Lorentz factor of a front
relative to the co-moving frame. For a flow with a large number $N_c$ of sheets, the average number of scattering before escaping is 
\begin{eqnarray}\label{equ4}
N_s \sim \log \frac{\epsilon_{cl}(r_b)}{\epsilon_0}/\log \frac{\gamma_*^2N_c}{N_c-1},\nonumber
\end{eqnarray}
where $\epsilon_{cl}$ is the local confinement energy limit. $N_s$ is typically of order $10$ for an initial energy $\epsilon_0$ between $1$ and $10^6$~GeV, as checked in numerical simulations.
Thus, the average energy gain by this secondary process is $<$$G$$>$ $\simeq\gamma_*^{2 N_s}\sim10^6$,
and the UHECR energy range can be achieved.
Then, UHECRs would produce a powerful synchrotron emission up to a few TeV for an observer.
The ratio of the synchrotron energy emitted by cosmic rays to that emitted by relativistic electrons is
\begin{eqnarray}\label{equ5}
\frac{E_{syn}^{cr}}{E_{syn}^{re}} = \frac{N_{cr}}{N_{re}}\frac{\epsilon_{max}^{cr}}{\epsilon_{max}^{re}} \nonumber
\left(\frac{m_e}{m_p}\right)^2 \sim 0.25.
\end{eqnarray}
This is an important amount of energy, but the number of counts could be poor because of the high energy of the emitted photons.
The number of high-energy photons emitted by the cosmic rays above some energy $\epsilon_{\gamma}$ can be estimated from the
energy $E_r$ received on the detector as electron synchrotron emission:
\begin{eqnarray}\label{equ6}
N_\mathrm{counts}(>\epsilon_{\gamma}) \simeq \frac{E_{syn}^{cr}}{E_{syn}^{re}} \frac{E_r}{\sqrt{\epsilon_{\gamma,min}.\epsilon_{\gamma}}},\nonumber
\end{eqnarray}
where $\epsilon_{\gamma,min}\sim 1$~GeV and $\epsilon_{\gamma}$ is in the GeV--TeV range.
Using\- typical va\-lues for the Band's parameters (i.e. $\alpha_\mathrm{Band}$$=$$-1$, $\beta_\mathrm{Band}$$=$$-2.25$ and
$E_\mathrm{peak}$$=$$200$~keV) and using a burst duration of $\sim$$20$~s, we expect GLAST--LAT to observe between 10 and 200 photons above $1$~GeV.
Such observation would confirm the UHECR production in GRBs.

At even higher energies ($\sim$$100$~GeV), the ability of Atmospheric Cerenkov Telescopes to detect this emission, which occurs in the early
stages of internal shocks, depends strongly on their slewing fastness.
Recent rapid observations of GRBs using the MAGIC telescope~\cite{Albert}, performed $1$ to $\sim$$10$~min after the burst, are encouraging.
However, a detection remains very difficult at these energies, since fluxes are strongly attenuated by the interaction of gamma rays 
(via pair-production) with photons from the extragalactic background light.


\section{Conclusions}
Future observations of the GRB prompt emission with GLAST should provide new estimates of the magnetic field parameters.
In particular the index of the turbulence spectrum should be determined by a kind of Amati's law and the existing data already favors a Kolmogorov
spectrum.
The level of the turbulence should be constrained.
Whatever its level, UHECRs should be generated by GRBs in the frame of the multi-fireball model through multiple scattering off magnetic fronts.
Their synchrotron emission in the GeV range should be a clear signature easily detectable by GLAST.

\bibliographystyle{plain}
\bibliography{icrc1079}

\end{document}